\documentclass{article}

\usepackage{spconf,amsmath,graphicx}
\ninept
\usepackage{bm}
\usepackage{amsthm}
\usepackage{amsfonts}
\usepackage{algorithm}
\usepackage{algpseudocode}

\title{An Improved DOA Estimation Algorithm for Circular and Non-Circular Signals with High Resolution}
\name{Liangtian Wan,  Member, IEEE, and Lihua Xie, Fellow, IEEE}
\address{ School of Electrical and Electronic Engineering\\
 Nanyang Technological University, 639798, Singapore}
\begin{document}

%
\maketitle
\begin{abstract}
In this paper, an improved direction-of-arrival (DOA) estimation algorithm for circular and non-circular signals is proposed. Most state-of-the-art algorithms only deal with the DOA estimation problem for the maximal non-circularity rated and circular signals. However, common non-circularity rated signals are not taken into consideration. The proposed algorithm  can estimates not only the maximal non-circularity rated and circular signals, but also the common non-circularity rated signals. Based on the property of the non-circularity phase and rate, the incident signals can be divided into three types as mentioned above, which can be estimated separately. The interrelationship among these signals can be reduced significantly, which means the resolution performance among different types of signals is improved. Simulation results illustrate the effectiveness of the proposed method.
\end{abstract}
\section{Introduction}
Direction-of-arrival (DOA) estimation has been one of the central problems in radar, sonar, navigation, geophysics, and acoustic tracking, which plays an important role in array signal processing \cite{Krim1996Two}. A wide variety of high-resolution narrowband DOA estimation algorithms have been proposed and analyzed in the past few decades \cite{Schmidt1986Multiple,Roy1989ESPRIT,Malioutov2005sparse,Hyder2010Direction,Nu2013DOA}. However, the DOA estimation methods above are only applicable to estimate circular signals, the DOA estimation performance of non-circular signals cannot be improved by using these methods, since they did not exploit the non-circularity property of non-circular signals.

The non-circular signals have been applied in various of communication systems. For example, amplitude modulated (AM) and binary phase shift keying (BPSK) modulated signals are widely used in satellite systems. Thus some algorithms \cite{Charge2001root,Haardt2004root} have been proposed for DOA estimation of non-circular signals. Existing works have also addressed the problem of how to increase the number of detectable sources and improve the estimation accuracy \cite{Liu2008Extended,Liu2012Direction,Steinwandt2014R}. However, these algorithms cannot deal with the situation where circular and non-circular signals coexist. In \cite{Abeida2006MUSIC}, the MUSIC-like algorithms have been extended to non-circular signals. However, as pointed out in \cite{Wan2015DOA}, the method in  \cite{Abeida2006MUSIC} has the drawback that when the maximal non-circularity rated, common non-circularity rated, and circular signals coexist. In the case of low SNR, small snapshot numbers, and large source numbers, more than one spectrum peakings may appear in the neighborhoods of DOAs of common non-circularity rated and circular signals in a certain probability. The method proposed in \cite{Gao2008MUSIC} estimates non-circular and circular signals simultaneously. However, when two types of signals are too close, this method cannot resolve them. The method proposed in \cite{Liu2012MUSIC} estimates non-circular and circular signals separately. Its estimation accuracy is lower than that of the MUSIC method in large DOA separation. Moreover, these algorithms \cite{Gao2008MUSIC,Liu2012MUSIC} cannot estimate the DOAs of the common non-circularity rated signals with high accuracy.

In this paper, we propose a novel DOA estimation algorithm for circular and non-circular signals with high resolution. Different types of signals are estimated separately. The preliminary DOAs for the maximal and common non-circularity rated signals are estimated first. Based on a new extended noise subspace, the DOAs for the common non-circularity rated signals are estimated with high accuracy. Then the effects of these signals on extended covariance matrix are eliminated by spatial differencing. Based on the previous work of the author in \cite{Wan2015DOA}, the DOAs for circular signals are estimated with the extended received data vector. The DOAs for the maximal non-circularity rated signals are estimated finally.

\section{PROBLEM FORMULATION}
Assume that $q$ narrowband plane wave signals are impinging on an $M$-element uniform linear array (ULA) with interelement spacing $d$ from $\bm{\theta} {\rm{ = }}\left[ {{\theta _1}, \ldots ,{\theta _q}} \right]$.
The output of $M$ sensors can be expressed as
\begin{equation}
{\bf{x}}(t) = {\bf{A}({\bm{\theta}})\mathbf{s}}(t) + {\bf{n}}(t),t = 1, \ldots ,N
\end{equation}
where ${\bf{s}}(t) \in {\mathbb{C}^{q \times 1}}$ represents the signal waveforms; ${\bf{n}}(t) \in {\mathbb{C}^{M \times 1}}$ is the white circular complex
Gaussian noise with zero-mean and variance ${\sigma ^2}$; $N$ is the number of available snapshots; ${\bf{A}({\bm{\theta}})} \in {\mathbb{C}^{M \times q}}$ is the manifold matrix with the form
\begin{equation}
{\bf{A}}\left( \bm{\theta} \right) = \left[ {{{\bf{a}}_1}\left( \theta  \right), \ldots ,{{\bf{a}}_q}\left( \theta  \right)} \right],
\end{equation}
and the steering vectoring of the $i$th $(i = 1,2, \ldots ,q)$ signal can be expressed as
\begin{equation}
{{\bf{a}}}\left( {{\theta _i}} \right) = {\left[ {1,{e^{j\tau \left( {{\theta _i}} \right)}}, \ldots ,{e^{j\left( {M - 1} \right)\tau \left( {{\theta _i}} \right)}}} \right]^T}\in {\mathbb{C}^{M \times 1}}
\end{equation}
where $\tau \left( {{\theta _i}} \right) = {{2\pi d\sin {\theta _i}} \mathord{\left/
 {\vphantom {{2\pi d\sin {\theta _i}} \lambda }} \right.
 \kern-\nulldelimiterspace} \lambda }$, $\lambda $ denotes the signal wave length. The covariance matrix $\mathbf{R}$ of ${\bf{x}}(t)$ can be expressed as
 \begin{equation}
\mathbf{R} = E\left\{ {\mathbf{x}\left( t \right){\mathbf{x}^H}\left( t \right)} \right\} = \mathbf{A}{\mathbf{R}_S}{\mathbf{A}^H} + \sigma _n^2{\mathbf{I}_M},
\end{equation}
where ${{\bf{R}}_S}{\rm{ = }}{\rm{diag}}\left\{ {\sigma _1^2, \ldots ,\sigma _q^2} \right\}$ is the signal covariance matrix, and $\sigma _i^2$ is the power of the $i$th signal.

For non-circular signal $s$, it holds that \cite{Abeida2006MUSIC}
\begin{equation}
{\mathop{\rm E}\nolimits} \left[ {s(t)s(t)} \right] = \rho {e ^{j\beta }}{\mathop{\rm E}\nolimits} \left[ {s(t){s^*}(t)} \right],
\end{equation}
in which $\beta $ and  $\rho$ are the non-circularity phase and rate, respectively. $\rho  = 1$ and $0 < \rho  < 1$ stand for the maximal and common non-circularity rated signal, respectively.

Assume that the signal vector ${\bf{s}}(t)$ contains $q$ non-circular signals,its unconjugated covariance
matrix is given by 
\begin{equation}
\begin{aligned}
 {{{\bf{R'}}}_S} &= E\left[ {{\bf{s}}(t){{\bf{s}}^T}(t)} \right]\: \\
 &= {\rm{diag}}\{ E\left[ {{s_1}(t){s_1}(t)} \right], \ldots ,E\left[ {{s_q}(t){s_q}(t)} \right]\} \: \\
& = {\rm{diag}}\{ {\rho _1}{e^{j{\beta _1}}}\sigma _1^2, \ldots ,{\rho _q}{e^{j{\beta _q}}}\sigma _q^2\} \buildrel \Delta \over = {\bf{PB}}{{\bf{R}}_S}, \\
\end{aligned}
\end{equation}
where $\mathbf{P}$ is a diagonal matrix, whose diagonal entries are the non-circularity rates of the $q$ signals and is defined as ${\bf{{P}}} = {{\rm diag}} \{ {\rho _1},{\rho _2}, \ldots ,{\rho _q}\} $. ${\bf{{ B}}}$ is a diagonal matrix, whose diagonal entries are the non-circularity phases of the $q$ signals and is defined as ${\bf{{B}}} = {{\rm diag}} \{ \exp \left( {j {\beta _1}} \right),\exp \left( {j {\beta _2}} \right), \ldots ,\exp \left( {j {\beta _q}} \right)\} $. When the $q$ signals are all circular signals, the non-circularity rate matrix satisfies ${{\bf{R'}}_S } = {\bf{0}}$.

As mentioned in \cite{Gao2008MUSIC}, it is more realistic that some users send non-circular symbols while other users still send circular symbols. Thus assume that the number of the maximal non-circularity rated, common non-circularity rated and circular signals are ${q_{ncm }}$, ${q_{ncn }}$ and ${q_c}$, respectively, with ${q_{ncm }} + {q_{ncn }} + {q_c} = q$. We aim to estimate the DOAs of the $q$ circular and non-circular signals by exploiting the $N$ snapshots of the array output vector ${\bf{x}}(t)$.

\section{The Proposed Method}
Since the unconjugated covariance matrix of the non-circularity rated of circular signals and noise all equal to zero, the unconjugated covariance matrix ${\bf{R'}}$  can be written as
\begin{equation}
\begin{array}{ccccc}
 {\bf{R'}}  = {\bf{APB}}{{\bf{R}}_S }{{\bf{A}}^T } = \left[ {\begin{array}{*{20}{c}}
   {{{\bf{A}}_{ncm }}} & {{{\bf{A}}_{ncn }}}  \\
\end{array}} \right] \times  \\
 {\kern 1pt} {\kern 1pt} \left[ {\begin{array}{*{20}{c}}
   {{{\bf{B}}_{ncm }}{{\bf{R}}_{{S _{ncm }}}}} & \mathbf{0}  \\
   \mathbf{0} & {{{\bf{P}}_{ncn }}{{\bf{B}}_{ncn }}{{\bf{R}}_{{S _{ncn }}}}}  \\
\end{array}} \right]\left[ {\begin{array}{*{20}{c}}
   {{\bf{A}}_{ncm }^T }  \\
   {{\bf{A}}_{ncn }^T }  \\
\end{array}} \right] \\
 \;{\kern 1pt}  = {{\bf{A}}_{ncm }}{{\bf{B}}_{ncm }}{{\bf{R}}_{{S _{ncm }}}}{\bf{A}}_{ncm }^T  + {{\bf{A}}_{ncn }}{{\bf{P}}_{ncn }}{{\bf{B}}_{ncn }}{{\bf{R}}_{{S _{ncn }}}}{\bf{A}}_{ncn }^T  \\
 \end{array}
\end{equation}
where the subscripts ${\left(  \cdot  \right)_{ncm }}$ and ${\left(  \cdot  \right)_{ncn }}$ stand for the matrix or vector corresponding to the maximal and common non-circularity rated signals, respectively. We construct a new matrix
\begin{equation}
\begin{array}{l}
 {\bf{R'}}{{{\bf{R'}}}^H } = {\bf{APB}}{{\bf{R}}_S }{{\bf{A}}^T }{{\bf{A}}^*}{\left( {{\bf{PB}}{{\bf{R}}_S }} \right)^H }{{\bf{A}}^H } \\
 {\kern 30pt}  \buildrel \Delta \over = {\bf{A}}{{\tilde{\bf{ R}}}_S }{{\bf{A}}^H }. \\
 \end{array}
\end{equation}
It can be known that ${{\rm rank}} \left\{ {{\bf{R'}}} \right\} = {{\rm rank}} \left\{ {{\bf{R'}}{{{\bf{R'}}}^H }} \right\}= {q_{ncm }} + {q_{ncn }} \le M - 1$. i.e., the matrix ${\bf{R'}}$ and ${\bf{R'}}{{\bf{R'}}^H }$ are not full rank matrices. Taking singular value decomposition (SVD) of ${\bf{R'}}{{\bf{R'}}^H }$, we have
\begin{equation}
\begin{array}{l}
 {\bf{R'}}{{{\bf{R'}}}^H } = {{\bf{Q}}_1}{{\bf{\Lambda }}^2}{\bf{Q}}_2^H  \\
 {\kern 29pt}  = \left[ {\begin{array}{*{20}{c}}
   {{{\bf{Q}}_{S 1}}} & {{{\bf{Q}}_{N 1}}}  \\
\end{array}} \right]\left[ {\begin{array}{*{20}{c}}
   {{\bf{\Lambda }}_S ^2} & {}  \\
   {} & {\bf{0}}  \\
\end{array}} \right]\left[ {\begin{array}{*{20}{c}}
   {{\bf{Q}}_{S 1}^H }  \\
   {{\bf{Q}}_{N 1}^H }  \\
\end{array}} \right] \\
 \end{array}
\end{equation}

where ${{\bf{Q}}_1}$ and ${{\bf{Q}}_2}$ stand for the left and right singular eigen-matrices of the matrix ${\bf{R'}}$, respectively. ${{\bf{\Lambda }}_S }$ is a diagonal matrix whose diagonal entries are constructed by ${q_{{\mathop{\rm ncm}\nolimits} }} + {q_{{\mathop{\rm ncn}\nolimits} }}$ non-zero singular values.

Based on traditional MUSIC algorithm, the spatial spectrum function ${f_{nc }}\left( \theta  \right)$ corresponding to the maximal and common non-circularity rated signals can be constructed as
\begin{equation}
{f_{nc }}(\theta ) = {{\bf{a}}^H }(\theta ){{\bf{Q}}_{N 1}}{\bf{Q}}_{N 1}^H {\bf{a}}(\theta ).
\end{equation}

\subsection{DOA Estimation for Common Non-Circularity Rated Signals}
However, the estimation accuracy is not improved since only the unconjugated covariance matrix is used. In this subsection, the conjugated and unconjugated covariance matrices are applied to improve the DOA estimation accuracy of the common non-circularity signals. The extended received data vector ${\bf{y}}(t)$ is constructed as
\begin{equation}
{\bf{y}}(t) = \left[ {\begin{array}{*{20}{c}}
   {{\bf{x}}(t)}  \\
   {{{\bf{x}}^*}(t)}  \\
\end{array}} \right].
\end{equation}
Then the eigen-value decomposition (EVD) of the covariance matrix ${{\bf{R}}_y}$ of ${\bf{y}}(t)$ is performed as
\begin{equation}
\begin{array}{ccccc}
 {{\bf{R}}_y}= {{\bf{U}}_S }{{\bf{\Sigma }}_S }{\bf{U}}_S ^H  + {{\bf{U}}_N }{{\bf{\Sigma }}_N }{\bf{U}}_N ^H  \\
 {\kern 10pt} \; = {{\bf{U}}_S }{{\bf{\Sigma }}_S }{\bf{U}}_S ^H  + \sigma _n ^2{{\bf{U}}_N }{\bf{U}}_N ^H.  \\
 \end{array}
\end{equation}
where the column vectors of $\mathbf{U}_S$ and $\mathbf{U}_N$ are constructed by the  eigenvectors corresponding to $q$ large eigenvalues and $2M-q$ small eigenvalues, respectively. The entries of diagonal matrices $\mathbf{\Sigma} _S$ and $\mathbf{\Sigma} _N$ are constructed by $q$ large singular values and $2M-q$ small singular values, respectively.

According to the structure of  ${\bf{y}}(t)$ in (11), the noise subspace ${{\bf{U}}_N }$ is partitioned into two block matrices
\begin{equation}
{{\bf{U}}_N } = \left[ {\begin{array}{*{20}{c}}
   {{{\bf{U}}_{N 1}}}  \\
   {{{\bf{U}}_{N 2}}}  \\
\end{array}} \right].
\end{equation}

According to \cite{Wan2015DOA}, for the common non-circularity rated and circular signals, the space spanned by the corresponding steering vectors is orthogonal to the block matrix ${{\bf{U}}_{N 1}}$. Thus a new matrix ${\bf{W}}$ is constructed
\begin{equation}
{\bf{W}} = \left[ {\begin{array}{*{20}{c}}
   {{{\bf{Q}}_{N 1}}} & {{{\bf{U}}_{N 1}}}  \\
\end{array}} \right],
\end{equation}
which is orthogonal to the common non-circularity rated signals with ${\bf{A}}_{ncn }^H {\bf{W}} = {\bf{0}}$. In order to improve the estimation accuracy, the column space of the matrix ${\bf{W}}$ needs to be combined, and the unit orthogonalization is taken for the columns. Since ${{\bf{Q}}_{N 1}}\in {\mathbb{C}^{M \times \left( {M - {q_{ncm }} - {q_{ncn }}} \right)}}$ and ${{\bf{U}}_{N 1}}\in {\mathbb{C}^{M \times \left( {2M - {q_{ncm }} - 2{q_{ncn }} - 2{q_c}} \right)}}$, we have ${\bf{W}}\in {\mathbb{C}^{M \times \left( {3M - 2{q_{ncm }} - 3{q_{ncn }} - 2{q_c}} \right)}}$. The SVD is taken on ${\bf{W}}$, i.e.,
\begin{equation}
{\bf{W}} = {{\bf{W}}_1}{{\bf{\Lambda }}_w}{\bf{W}}_2^H,
\end{equation}
where ${{\bf{W}}_2}\in {\mathbb{C}^{\left( {3M - 2{q_{ncm }} - 3{q_{ncn }} - 2{q_c}} \right) \times \left( {3M - 2{q_{ncm }} - 3{q_{ncn }} - 2{q_c}} \right)}}$ and ${{\bf{W}}_1}\in {\mathbb{C}^{M \times M}} $ are the right and left singular eigen-matrices of ${\bf{W}}$, respectively; ${{\bf{\Lambda }}_w}$ is a diagonal matrix whose diagonal entries are constructed by the non-zero singular values of ${\bf{W}}$. Then ${{\bf{W}}_1}$ can be partitioned into block matrices as
\begin{equation}
{{\bf{W}}_1} = \left[ {\begin{array}{*{20}{c}}
   {{{\bf{W}}_{11}}} & {{{\bf{W}}_{12}}}  \\
\end{array}} \right],
\end{equation}
where ${{\bf{W}}_{11}}\in {\mathbb{C}^{M \times w}}$ is orthogonal to the manifold matrix of the common non-circularity rated signals. i.e.,
\begin{equation}
{{\bf{A}}_{ncn }}^H {{\bf{W}}_{11}} = {\bf{0}},
\end{equation}
and $w$ satisfies
\begin{equation}
w = \min \left\{ {3M - 2{q_{ncm }} - 3{q_{ncn }} - 2{q_c},M - {q_{ncn }}} \right\}.
\end{equation}
The spatial spectrum function for the common non-circularity rated signals can be constructed as
\begin{equation}
{f_{ncn }}(\theta ) = {{\bf{a}}^H }(\theta ){{\bf{W}}_{11}}{\bf{W}}_{11}^H {\bf{a}}(\theta ).
\end{equation}

\subsection{DOA Estimation for Circular Signals}
Based on the DOA estimation of the common non-circularity rated signal, its manifold matrix can be reconstructed as ${\hat{\bf{ A}}_{ncn }} = {\bf{A}}({\hat \theta _{ncn }})$. It can be seen that the low accuracy DOA estimate of (10) contains the maximal and common non-circularity rated signals. For the common non-circularity rated signals,  the DOA estimation of (10) can be replaced by (19). The low accuracy manifold matrix estimation of  ${q_{ncm }}$  maximal non-circularity rated signals can be obtained ${\dot{\hat{\bf{ { A}}}}}_{ncm } = {\bf{A}}({\dot {\hat \theta }_{ncm }})$, where ${\dot {\hat \theta }_{ncm }}$ stands for the low accuracy DOA estimation of the maximal non-circularity rated signal. Then the manifold array ${{\bf{\hat A}}_{nc }}$ of the maximal and common non-circularity rated signals can be reconstructed as
\begin{equation}
{\hat{\bf{ A}}_{nc }} = \left[ {\begin{array}{*{20}{c}}
   {{{\dot{\hat{\bf{ { A}}}}}_{ncm }}} & {{{\hat{\bf{ A}}}_{ncn }}}  \\
\end{array}} \right].
\end{equation}
Based on the property of pseudo inverse matrix \cite{Liu2012MUSIC}, (7) can be rewritten as
\begin{equation}
\begin{array}{ccccc}
 {\left( {{{\hat{\bf{ A}}}_{nc }}} \right)^\dag }{\bf{R'}} {\left( {\hat{\bf{ A}}_{nc }^T } \right)^\dag } = {\bf{PB}}{{\bf{R}}_S } \\
   = \left[ {\begin{array}{*{20}{c}}
   {{{\bf{B}}_{ncm }}{{\bf{R}}_{{S _{ncm }}}}} & \mathbf{0}  \\
   \mathbf{0} & {{{\bf{P}}_{ncn }}{{\bf{B}}_{ncn }}{{\bf{R}}_{{S _{ncn }}}}}  \\
\end{array}} \right]. \\
 \end{array}
\end{equation}
Then the matrix in (18) can be partitioned into block matrices as follows
\begin{equation}
{\left( {{{\hat{\bf{ A}}}_{nc }}} \right)^\dag }{\bf{R'}} {\left( {\hat{\bf{ A}}_{nc }^T } \right)^\dag } = \left[ {\begin{array}{*{20}{c}}
   {{{\bf{R}}_1}} & {{{\bf{R}}_2}}  \\
   {{{\bf{R}}_3}} & {{{\bf{R}}_4}}  \\
\end{array}} \right],
\end{equation}
where ${{\bf{R}}_1}\in {\mathbb{C}^{{q_{ncm }} \times {q_{ncm }}}}$ and ${{\bf{R}}_2}\in {\mathbb{C}^{{q_{ncn }} \times {q_{ncn }}}}$.  Then the unconjugated covariance matrix of the common non-circularity rated signals can be estimated as
\begin{equation}
{{\bf{\hat R'}}_{{S _{ncn }}}} = {{\bf{R}}_4}.
\end{equation}

In modern wireless communication systems, the modulation modes of the signals can be detected. The non-circularity rated of the non-circular signals can be confirmed uniquely, i.e., the non-circularity rated matrix ${\hat{\bf{ P}}_{ncn }}$ of the non-circular signals is known as a prior information. Then the covariance matrix of the common non-circularity rated signals can be estimated as
\begin{equation}
{\left( {{{{\bf{\hat R}}}_{{{\mathop{\rm S}\nolimits} _{{\mathop{\rm ncn}\nolimits} }}}}} \right)_{ii}} = \left\| {{{\left( {{\bf{\hat P}}_{{\mathop{\rm ncn}\nolimits} }^{ - 1}{{\bf{R}}_4}} \right)}_{ii}}} \right\|,
\end{equation}
where ${\left(  \cdot  \right)_{ii}}$ stands for the  $i$th diagonal entry of a diagonal matrix. Based on (20), the conjugated and unconjugated covariance matrices ${\hat{\bf{ R}}_{ncn }}$ and ${\hat{\bf{ R'}}_{ncn }}$ for the common non-circularity rated signals can be respectively estimated as
\begin{equation}
{\hat{\bf{ R}}_{ncn }} = {\hat{\bf{ A}}_{ncn }}\left\| {{{\left( {\hat{\bf{ P}}_{ncn }^{ - 1}{{\bf{R}}_4}} \right)}_{ii}}} \right\|\hat{\bf{ A}}_{ncn }^H,
\end{equation}
\begin{equation}
{\hat{\bf{ R'}}_{ncn }} = {\hat{\bf{ A}}_{ncn }}{{\bf{R}}_4}\hat{\bf{ A}}_{ncn }^T.
\end{equation}

Since the DOAs of the common non-circularity rated signals have already been estimated, the information of the common non-circularity rated signals can be eliminated from the covariance matrix ${{\bf{R}}_y}$. A new matrix ${\tilde {\bf{R}}_{ncn }}$ can be constructed as
\begin{equation}
{\tilde{\bf{ R}}_{ncn }} = \left[ {\begin{array}{*{20}{c}}
   {{{\hat{\bf{ R}}}_{ncn }}} & {{{\hat{\bf{ R'}}}_{ncn }}}  \\
   {{{\left( {{{\hat{\bf{ R'}}}_{ncn }}} \right)}^*}} & {{{\left( {{{\hat{\bf{ R}}}_{ncn }}} \right)}^*}}  \\
\end{array}} \right].
\end{equation}
Then a differencing matrix ${{\bf{R}}_{y - ncn }}$ can be defined as
\begin{equation}
{{\bf{R}}_{y - ncn }} = {{\bf{R}}_y} - {\tilde{\bf{ R}}_{ncn }}.
\end{equation}

According to the analysis in \cite{Wan2015DOA}, for the common non-circularity rated and circular signals, the corresponding steering vectors satisfy
\begin{equation}
\left\{ \begin{array}{l}
 {\mathbf{a}^H}(\theta ){\mathbf{U}_{N 1}} + {e ^{j \phi }}{\mathbf{a}^T}(\theta ){\mathbf{U}_{N 2}} = \mathbf{0}, \\
  - {\mathbf{a}^H}(\theta ){\mathbf{U}_{N 1}} + {e ^{j \phi }}{\mathbf{a}^T}(\theta ){\mathbf{U}_{N 2}} = \mathbf{0}, \\
 \end{array} \right.
\end{equation}
i.e.,
\begin{equation}
{\mathbf{a}^H}(\theta ){\mathbf{U}_{N 1}} = \mathbf{0},{\mathbf{a}^T}(\theta ){\mathbf{U}_{N 2}} = \mathbf{0}.
\end{equation}
It can be seen that the two equations in (30) have the same information based on ${{\bf{U}}_{N 2}} = \mathbf{U}_{N 1}^*\bf{\Delta} $ \cite{Abeida2006MUSIC}. For the circular signals, the spectrum estimation function can be expressed as
\begin{equation}
f(\theta )= {\left\| {{{\bf{a}}^H }(\theta ){{\bf{U}}_{N 1}}} \right\|_{\mathop{\rm F}\nolimits} } = {{\bf{a}}^H }(\theta ){{\bf{U}}_{N 1}}{\bf{U}}_{N 1}^H {\bf{a}}(\theta ).
\end{equation}
The DOAs of the circular signals can be obtained by searching $q_c$ minimum values of (31). The DOA estimation accuracy of the circular signals should be improved since the covariance of the extended received data vector is used.

\subsection{DOA Estimation for Maximal Non-Circularity Rated Signals}
It should be known that the  extended received data vector has to be used in order to improve the DOA estimation performance of the maximal non-circularity rated signals.
For the  maximal non-circularity rated signals, it should be noted that the extended steering vectors satisfies
\begin{equation}
{\left[ {\begin{array}{*{20}{c}}
   {{{\bf{a}}_i}(\theta )}  \\
   {{e ^{ - j{\beta _i}}}{\bf{a}}_i^*(\theta )}  \\
\end{array}} \right]^H }\left[ {\begin{array}{*{20}{c}}
   {{{{\bf{\bar U}}}_{N 1}}}  \\
   {{{{\bf{\bar U}}}_{N 2}}}  \\
\end{array}} \right] = {\bf{0}},\quad i = 1,2, \ldots {q_{ncm }}.
\end{equation}

According to the analysis in \cite{Wan2015DOA}, the spectrum estimation function of the maximal non-circularity rated signals can be expressed as
\begin{equation}
f(\theta ) = {{\bf{a}}^H }(\theta ){{\bf{U}}_{N 1}}{\bf{U}}_{N 1}^H {\bf{a}}(\theta ) - \left| {{{\bf{a}}^T }(\theta ){{\bf{U}}_{N 2}}{\bf{U}}_{N 1}^H {\bf{a}}(\theta )} \right|.
\end{equation}
However, the DOAs of circular signals can make (33) achieve the minimum as well. In order to eliminate the effect of circular signals, some measures should be taken to solve this problem.  For the maximal and common non-circularity rated signals, it holds that ${{\bf{a}}^H }(\theta ){{\bf{Q}}_{N 1}} = {\bf{0}}$. However, the circular signals do not satisfy this relationship. Thus a new spectrum function for the maximal non-circularity rated signals can be constructed as
\begin{equation}
\begin{array}{l}
 {f_{ncm }}(\theta ) = {{\bf{a}}^H }(\theta )\left( {{{\bar {\bf{U}}}_{N 1}}\bar{\bf{ U}}_{N 1}^H  + {{\bf{Q}}_{N 1}}{\bf{Q}}_{N 1}^H } \right){\bf{a}}(\theta ) -  \\
 {\kern 42pt} {\kern 1pt} \left| {{{\bf{a}}^T }(\theta ){{\bar U}_{N 2}}\bar{\bf{ U}}_{N 1}^H {\bf{a}}(\theta )} \right|. \\
 \end{array}
\end{equation}

Thus, the DOA estimation for the common non-circularity rated signals, circular signals and  maximal non-circularity rated signals is completed finally. And the proposed algorithm called high resolution MUSIC (HRNC-MUSIC) algorithm has higher resolution than that of the DRNC-MUSIC and DRNC-MUSIC-C algorithm proposed in \cite{Wan2015DOA}.

\section{The Algorithm Analysis }
Based on the algorithms proposed in \cite{Wan2015DOA}, the HRNC-MUSIC algorithm exploits the characteristic of non-circularity further. When the incident signals co-exist the maximal non-circularity rated, common non-circularity rated and circular signals, the HRNC-MUSIC algorithm can estimate them in turn. The phenomenon of pseudo peakings appearing at the neighbourhoods of DOAs of the common non-circularity rated and circular signals in the algorithms proposed in \cite{Abeida2006MUSIC, Wan2015DOA} is avoided in the HRNC-MUSIC algorithm.

In order to validate the performance of the angular resolution of the proposed algorithm, an example is given as follows. Assume five narrowband farfield signals impinge on a ULA with five elements. The interelement spacing is $d = {\lambda  \mathord{\left/{\vphantom {\lambda  2}} \right.\kern-\nulldelimiterspace} 2}$, the signal-to-noise ratio (SNR) is fixed at 3dB, and the snapshot number is fixed at 500. The DOAs of two BPSK signals (non-circular ones with non-circularity rate of 1) are $35^\circ$ and $95^\circ$, respectively. The DOAs of two QPSK signals (circular ones with non-circularity rate of 0) are $40^\circ $ and $125^\circ$, respectively. The DOA of one UQPSK signal (the common non-circularity rated signal) is $135^\circ $. The spatial spectrum curves of DRNC-MUSIC algorithm and DRNC-MUSIC-C and HRNC-MUSIC algorithms are depicted in Fig. 1(a) and Fig. 1(b), respectively.
\begin{figure}[htb]
\begin{minipage}[b]{.48\linewidth}
  \centering
  \centerline{\includegraphics[width=4.0cm]{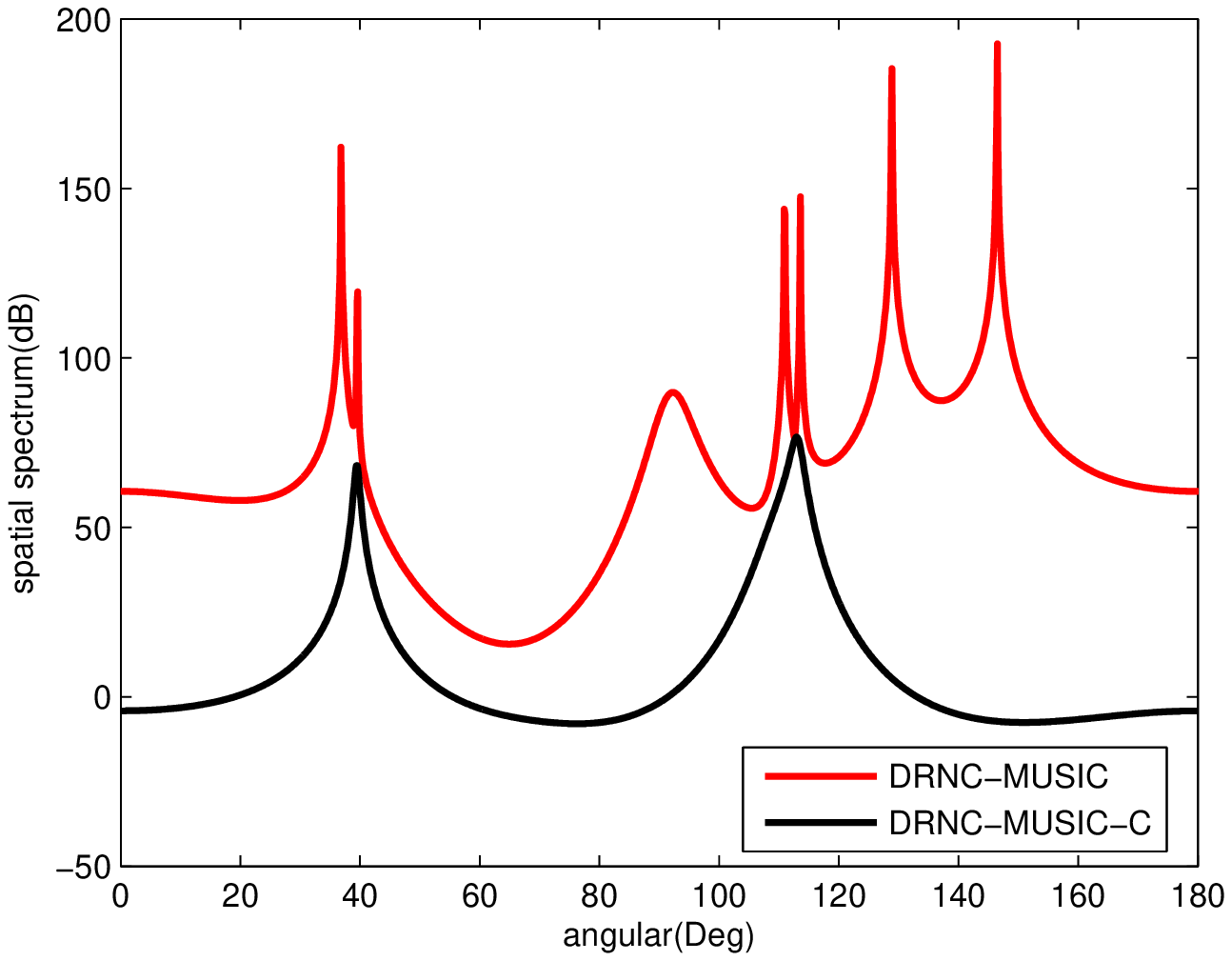}}
  \centerline{(a)  Algorithms in \cite{Wan2015DOA} }\medskip
\end{minipage}
\hfill
\begin{minipage}[b]{0.48\linewidth}
  \centering
  \centerline{\includegraphics[width=4.0cm]{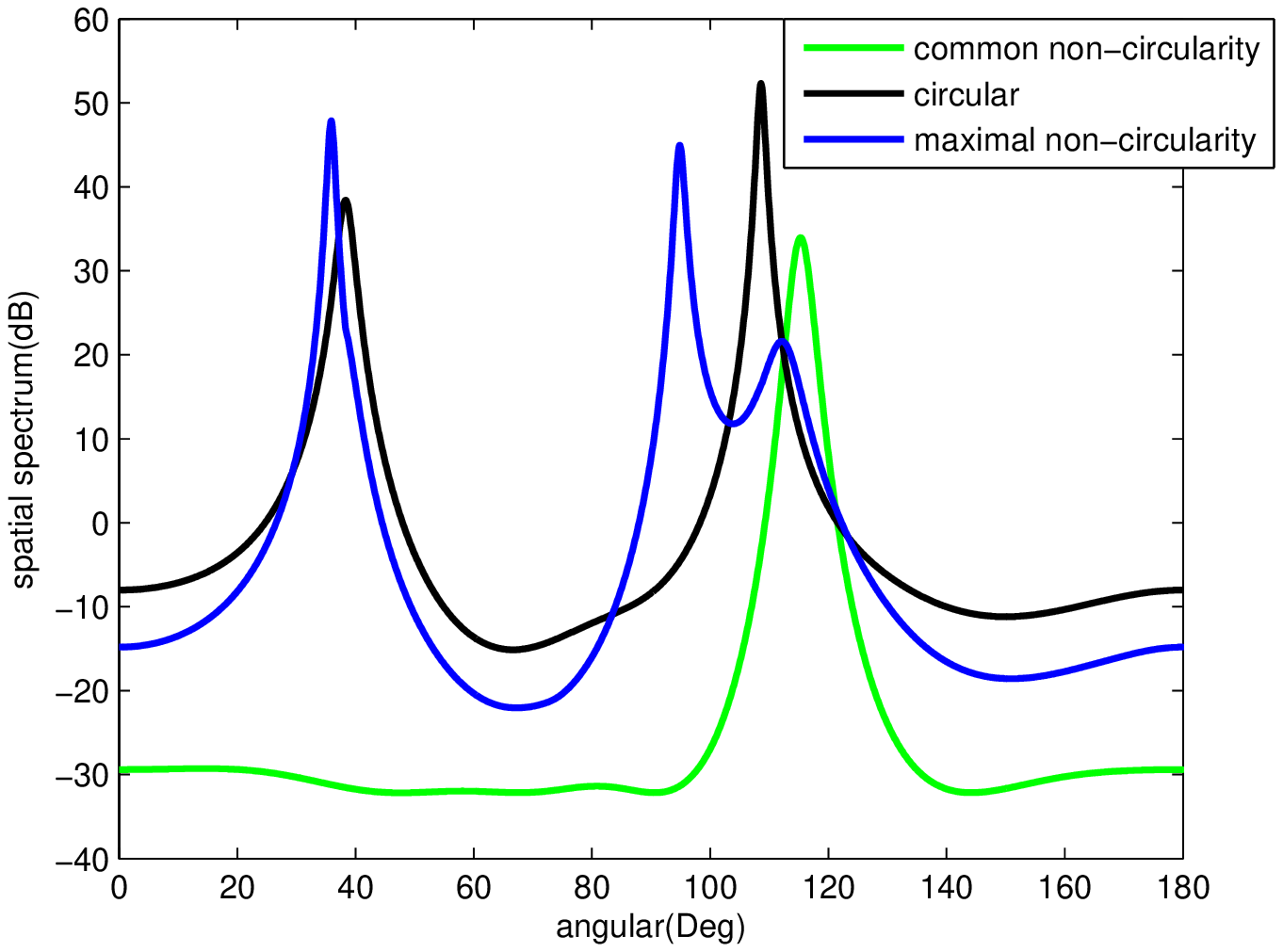}}
  \centerline{(b) HRNC-MUSIC}\medskip
\end{minipage}
\caption{The spatial spectrum of different algorithms.}
\label{fig:res}
\end{figure}

It can be seen from Fig. 1(a) that the pseudo peakings are obvious which are estimated by the DRNC-MUSIC algorithm. The DOAs of the maximal non-circularity rated signals are severely affected by these pseudo peakings. The DRNC-MUSIC-C algorithm can only estimate the common non-circularity rated and circular signals. When the angular distance of two types of signals are too close, the DRNC-MUSIC-C algorithm can not resolve them exactly. However, from Fig. 1(b), it can be seen that three spatial spectrums are utilized which are corresponding to the common non-circularity rated, circular and maximal non-circularity rated signals, respectively. Three types of signals are estimated in turn by the HRNC-MUSIC algorithm, the phenomenon of pseudo peakings is avoided, and the interrelationship between two different types of signals can be avoided as well.

\begin{figure}[htb]
\begin{minipage}[b]{.48\linewidth}
  \centering
  \centerline{\includegraphics[width=4.0cm]{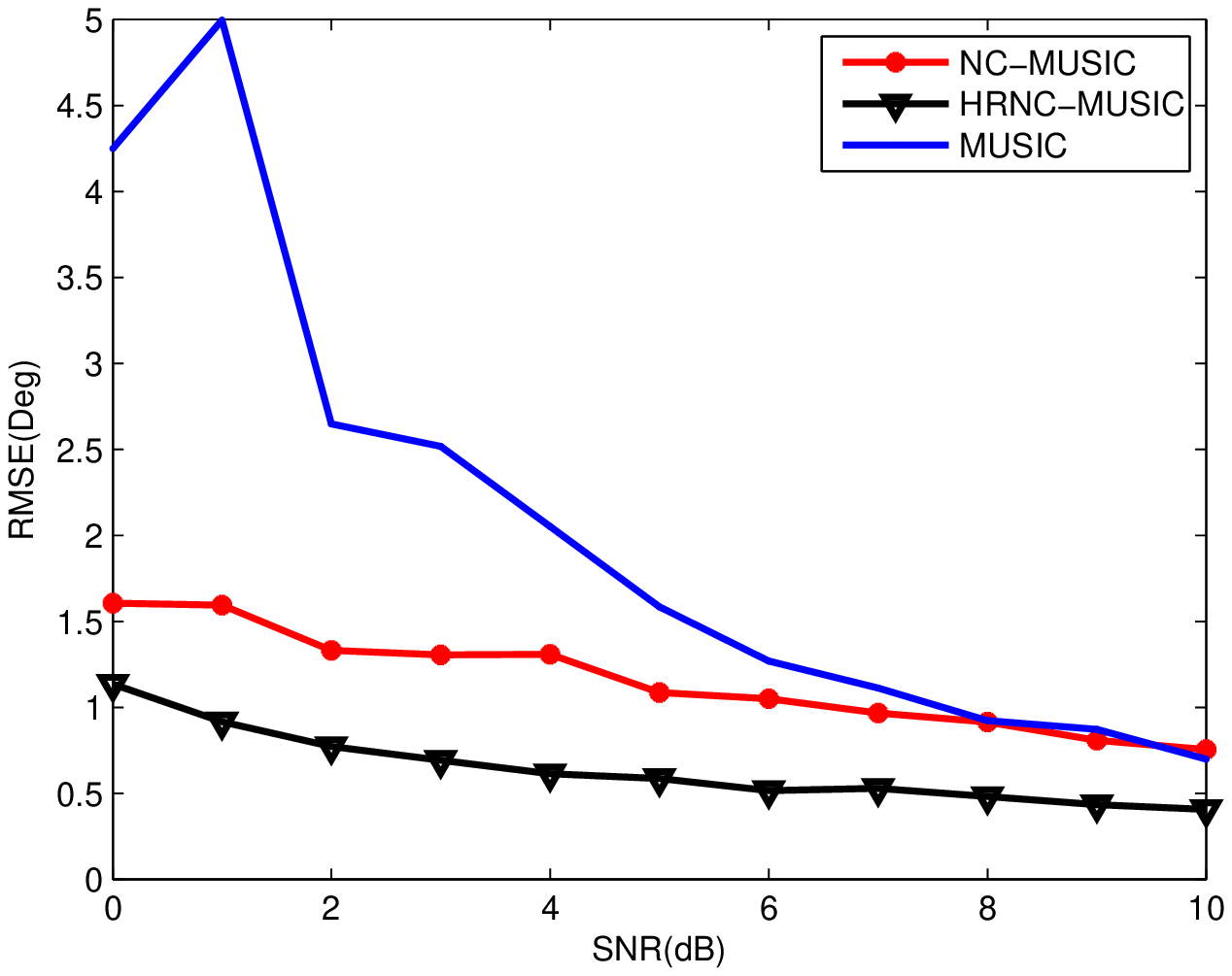}}
  \centerline{(a) RMSE  }\medskip
\end{minipage}
\hfill
\begin{minipage}[b]{0.48\linewidth}
  \centering
  \centerline{\includegraphics[width=4.0cm]{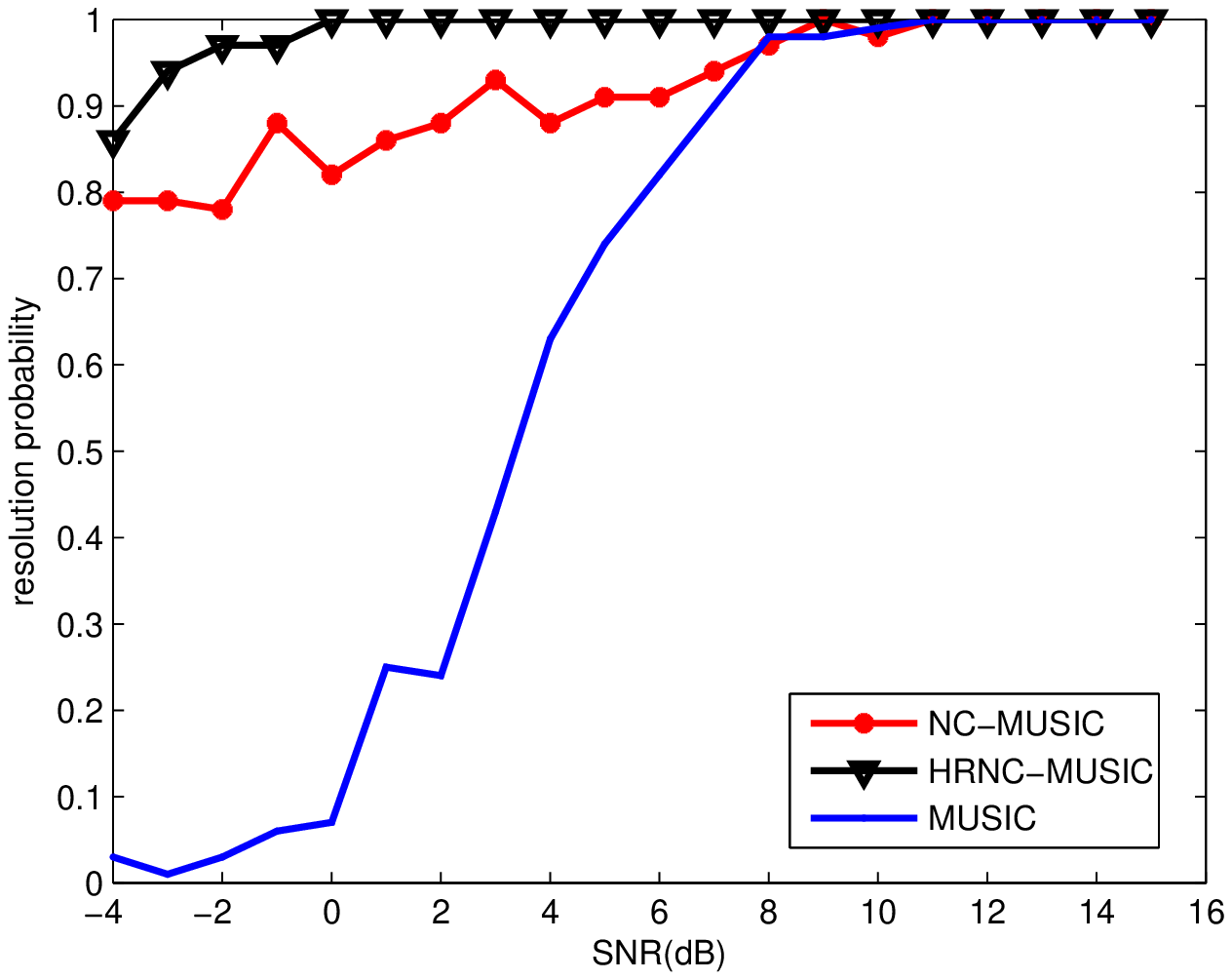}}
  \centerline{(b) Resolution probability}\medskip
\end{minipage}
\caption{The estimation performance versus SNR.}
\label{fig:re}
\end{figure}

\section{Simulation Results}
In the following examples, a ULA with interelement spacing $d = {\lambda  \mathord{\left/{\vphantom {\lambda  2}} \right.\kern-\nulldelimiterspace} 2}$ is employed. we compare the performance of the proposed (HRNC-MUSIC) algorithm, the MUSIC algorithm \cite{Schmidt1986Multiple}, and the NC-MUSIC algorithm in \cite{Abeida2006MUSIC} by simulations. All the results are averaged over $100$ Monte Carlo runs.

In the first example, we compare the estimation accuracy of different algorithms under the coexistence of both circular and non-circular sources. In order to test the estimation performance of HRNC-MUSIC algorithm, different types of signals with small angular distance are selected. We assume that four uncorrelated signals from $35^\circ $, $65^\circ $, $75^\circ $ and $85^\circ $, respectively, impinge on a six-element ULA, and consider three cases where there are two, one, and one sources, respectively. For the case with two sources, the source coming from $35^\circ $ and $65^\circ $ are supposed to send BPSK symbols with the non-circularity phases $10^\circ $ and  $20^\circ $, respectively; for the case with the former one source, the source from $75^\circ $ sends QPSK symbol, for the case with the latter one source, the source from $85^\circ$ sends QPSK symbol with the non-circularity phase  $40^\circ $ and the non-circularity rated $0.5$. The snapshot number is fixed at $500$. The root mean-square error (RMSE) is used to evaluate the performance of the algorithms.

The RMSEs versus the SNR of different algorithms are shown in Fig. 2(a). We see that the HRNC-MUSIC algorithm performs better than the traditional MUSIC and NC-MUSIC algorithm.   Moreover, it is noticed that the performance of the proposed method becomes better when SNR increases. This result is mainly due to that the HRNC-MUSIC algorithm estimates different types of signals separately. The effect of different types of signals is reduced, which makes the algorithm has a high estimation accuracy. However, another two algorithms estimate them simultaneously, the estimation accuracy is affected by the interrelationship among different types of signals.

The second example studies the resolution probability of different algorithms under different SNRs. The simulation condition is the same as the first example. The resolution probability versus the SNR of different algorithms are shown in Fig. 2(b). it can be seen that the resolution probability of different algorithms increases as the SNR increases. The resolution probability of the HRNC-MUSIC algorithm is higher than that of the NC-MUSIC and MUSIC algorithms. The estimation performance of the HRNC-MUSIC algorithm is stable at high SNR.


%
\section{Conclusions}
A novel DOA estimation algorithm is proposed under the coexistence of circular and non-circular signals.  The maximal non-circularity rated,  common non-circularity rated and circular signals can be estimated separately. The interrelationship among these signals can be reduced significantly, resulting in a higher resolution. The proposed algorithm performs better than traditional MUSIC and NC-MUSIC algorithm in small angular distance. In the future, we will focus on the DOA estimation algorithm under the coexistence of circular and non-circular signals with a low computational complexity.


\section{Acknowledgment}
The author would thank Dr. T. Zhu for his useful comments on the DOA estimation of the  maximal non-circularity rated signals.

\vfill\pagebreak

\bibliographystyle{IEEEbib}

\bibliography{reference}

\end{document}